\pdfoutput=1
\documentclass[11pt]{article} 

\usepackage[utf8]{inputenc} 

\usepackage{geometry} 
\usepackage{graphicx} 

\usepackage{booktabs} 
\usepackage{array} 
\usepackage{paralist} 
\usepackage{verbatim} 
\usepackage{subfig} 
\usepackage{graphicx}

\usepackage{color}
\usepackage{url}
 
\usepackage{fancyhdr} 
\usepackage{sectsty}
\allsectionsfont{\sffamily\mdseries\upshape} 

\usepackage[nottoc,notlof,notlot]{tocbibind} 
\usepackage[titles,subfigure]{tocloft} 

\title{Cortical thickness and functional networks modules by cortical lobes}
\author{Vesna Vuksanovi\'c\\\\
Aberdeen Biomedical Imaging Centre \\
              University of Aberdeen \\\\
            vesna.vuksanovic@abdn.ac.uk
}
         
\begin{document}
\maketitle

\section{Abstract}
This study aims to investigate topological organization of cortical thickness and functional networks by cortical lobes. First, I demonstrated modular organization of these networks by the cortical surface frontal, temporal, parietal and occipital divisions. Secondly, I mapped the overlapping edges of cortical thickness and functional networks for positive and negative correlations. Finally, I showed that overlapping positive edges map onto within-lobe cortical interactions and negative onto between-lobes interactions. \\
\section{Introduction}
Classical, localized characterization of human brain surface morphology has moved in recent years towards maps of inter-regional co-variations of gray matter volume or thickness measurements in structural magnetic resonance imaging data (sMRI) \cite{evans2013networks}. To make such cortical maps or brain networks (graphs) the edges (links) between the regions (nodes) are defined by the strength of correlation between regional volume or cortical thickness. The physiological mechanisms underlying the thickness correlation among cortical areas remain unclear. Neuroimaging findings propose  strong correlation of the cortical thickness measurements between regions that are directly connected via axonal connections \cite{lerch2006mapping,gong2012convergence}. However, only a fraction of variations in regional thickness correlations across the cortex has been explained by direct axonal links \cite{gong2012convergence}. In this context it is interesting to ask whether the regional thickness (morphological) co-variations also reflect functional associations (or parts of them). Recent studies suggest that functional network properties, such as the strengths or the number of correlations, can impact structural properties. The synchronization of neuronal activity in response to specific functional demands might induce synchronized plastic changes among related regions during brain development \cite{hyde2009effects} or degenerative diseases \cite{witiuk2014cognitive}. More pertinent to this study, such interplay between regional structural and functional properties may contribute to strong thickness correlations observed within the visual areas of the occipital lobe \cite{vuk2019cortical}. Here it is also worth noting that, while it has been observed that functional modules mirror the local brain anatomy a number of studies show deviations between these two suggesting many-to-one function-structure mapping \cite{park2013structural}.

The natural divisions of the cortical surface -- frontal, temporal, parietal and occipital -- have been defined by regional morphological characteristics, which are also known to support different functions \cite{kandel2000principles,mesulam1998sensation}. In this context, questions arise about the relationship between cortical surface inter-regional correlations, i.e., how do variations in functional and thickness correlations map onto each other? To what extent these cortical structural and functional correlations share similar topological organization? If topological measures of functional network organization are structurally relevant we might expect them to be impacted at the lobe level of the cortical surface organization. 

To address these questions I examined thickness and functional correlational networks based on two different neuroimaging modalities -- structural and functional MRI. The networks were represented by $68 \times 68$ correlation matrices, where functional and thickness network maps inter-regional correlations between brain activity and thickness measurements respectively. The main focus of the study is to shed light on intrinsic topological properties of the frontal, temporal, parietal and occipital divisions, which may impact structural and functional interactions across the cortical surface. To this end I studied how modular organization of the cortical surface divisions affect overlap of the cortical thickness and functional correlations across cortical regions. 
\section{Methods}
\label{sec:1}
The MRI data sets considered in this short communication are from a public database ($http://fcon_1000.projects.nitrc.org/$) provided by the Max Planck Institute (MPI) Leipzig. A study group included 37 (16M/21F) participants between 20 and 42 years. MRI data was acquired on 3 Tesla Magnetom Tim Trio scanner (Siemens, Erlangen, Germany) using a 32-channel head coil. T1-weighted images were acquired using a MPRAGE sequence (TR = 1.3 s; TE=3.46 ms; flip angle=10$\deg$; FOV=$256\times240 mm^2$; 176 sagittal slices; voxel size = $1\times1\times1.5 mm$). Functional MRI/EPI data were acquired on a 3T MRI scanner (Siemens Tim Trio) using TR=2.3 sec, TE=30ms, $3\times3$ in-plane resolution, 3 mm slice thickness, 1 mm gap between slices. Each scanning session was a task-absent ("resting state") scan lasting 7.6 minutes during which subjects were asked to fixate a fixation cross.

Processing the structural MRI data was done using FreeSurfer v5.3.0 (\url{https://surfer.nmr.mgh.harvard.edu/}) pipeline according to the procedure described in more details in \cite{vuk2019cortical}. Cortical thickness was measured on $N$ = 68 cortical surface regions segmented according to the Desikan-Keliany Atlas (DKA) \cite{desikan2006automated}. FSL v5.0.11 (\url{https://fsl.fmrib.ox.ac.uk/fsl/fslwiki/}) was used to process fMRI data and extract BOLD fMRI time-series on the same 68 cortical regions of the DKA parcellations. The fMRI data were processed using the FSL toolbox FEAT fMRI analysis, similarly to our previous work \cite{vuksanovic2014functional} and global signal regression was applied in order to reduce motion artifacts \cite{ciric2017benchmarking}.
\begin{figure}
\centering
\includegraphics[width=1\linewidth]{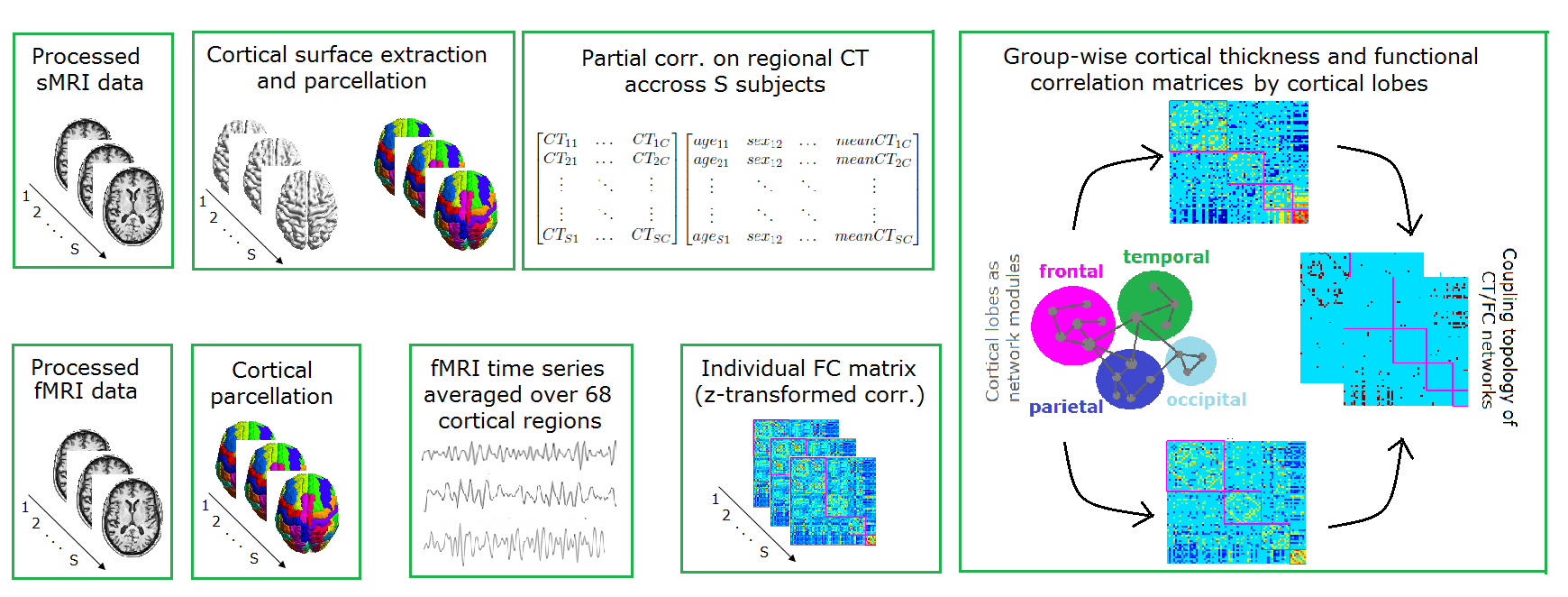}
\linespread{1}
\caption{Workflow diagram of the cortical networks' construction on processed structural and functional magnetic resonance images. (Upper panel) Structural magnetic resonance images (sMRI) are processed using FreeSurfer v5.3.0 pipeline: cortical surface was reconstructed and regional cortical thickness measures were averaged over N = 68 structures of the Desikan-Keliany Atlas (DKA). Structural correlation network was constructed on partial correlations between regional thickness across all participants (S = 37) while controlling for age, sex and mean CT. (Lower panel) Functional MRI (fMRI) were parcellated and time series were averaged over the same 68 DKA regions. Individual functional correlations (FC) matrix was constructed on pair-wise correlations between the fMRI time series, Fisher z-transformed, averaged over S subjects and transformed back to linear Pearson coefficient. Group-wise CT and FC matrices were reordered according to node affiliation with cortical lobes (frontal, temporal, parietal and occipital) and their coupling topology was assessed to capture positive and negative correlations maps.}  
\label{fig:workflow}
\end{figure}
\subsection{Brain Graph Construction}
Brain graphs considered in this study were constructed on correlations between either regional cortical thickness or BOLD activity measured by structural or functional MRI. In these correlational networks, a brain region represent a network node and a pair-wise correlation between the regional measurements represents network's edge/link or connection between the nodes. See Fig.~\ref{fig:workflow} that summarizes analysis pipeline and approach.
\subsubsection{Functional Correlations}
To obtain the FC matrices, time series of the 68 DKA regions were calculated by averaging the respective BOLD signals over all regional voxels. The correlation matrices were obtained for each subject in the study. Since voxels are in MNI space and a given voxel has approximately the same anatomical position in all subjects, the individual correlation matrices can be averaged across subjects. In detail, each matrix is first Fisher z-transformed and averaged across subjects and then transformed back into correlation coefficients again as described in previous studies \cite{vuksanovic2014functional,vuksanovic2015dynamic}. The resulting correlation matrix $f_{ij}, (i,j = 1,…,N)$, was used to create a group-wise functional correlation matrix between the regions of interest. See Fig.~\ref{fig:workflow} (lower panel) for analysis pipeline and approach.
\subsubsection{Cortical Thickness Correlations}
A group-wise structural network was constructed on inter-regional cortical thickness correlations across all study participants. The correlations were calculated between each pair of 68 regional thickness measurements while controlling for age, gender and mean CT, similarly to previously described methodology \cite{vuk2019cortical,gong2012convergence}. See Fig.~\ref{fig:workflow} (upper panel) for analysis pipeline and approach.
\subsection{Brain Graph Analysis}
\subsubsection{Network Density}
Network topological properties depend on the network thresholding. Threshold affects network density ($\kappa$), i.e., number of links relative to the total number of all possible links in the network \cite{WIJ10}. The threshold considered here was set to yield a fully connected network, i.e., one network component.  In addition, the threshold was applied to ensure that both networks CT and FC have the same density, i.e., to ensure that the networks are analyzed and compared across the same number of links. 
\subsubsection{Modularity by Lobes}
Each network was assessed at the scale of its lobar organization -- the frontal, temporal, parietal and occipital divisions of the cortical surface. Modularity index (Q), was calculated to determine whether cortical lobes as conventionally defined correspond to modules of the CT/FC network. This can be done by calculating the Q according to lobe (by employing vector of nodal affiliation with the particular lobe as initial community affiliation vector). The modularity index quantifies the observed fraction of within-module degree values relative to those expected if connections were randomly distributed across the network. In the context of this short communication, cortical lobes have been considered and analyzed as network modules. Since the constructed CT/FC network contains both positive and negative edge strengths, I used the asymmetric generalization of the modularity quality function introduced in Rubinov and Sporns \cite{RUB10}. Since the algorithm returns varying results on different runs, to control for the variance in the analysis, modularity index Q was calculated 100 times for each network, i.e., from running the same analysis 100 times on CT/FC network. In other words, index Q was estimated from 100 runs on CT/FT network and on 100 randomized (CT/FC each) versions of these networks. (See  section~\ref{sec:modlob}) for details.) For networks of similar size, it is accepted that a Q value above 0.3 is a good indicator of the existence of significant modules in a network \cite{clauset2004finding}. 
\section{Results}
\subsection{Network Density}
As described in the Methods, a single value for network density was obtained by considering percolation threshold for each network. This yielded a network density of $\kappa$ = 0.16, which includes both positive and negative edges. This choice of the threshold insures that each network (either CT or FC) has the same ($N = 362$) total number of positive and negative links. 
\label{sec:2}
\begin{figure}
\centering
\includegraphics[width=1\linewidth]{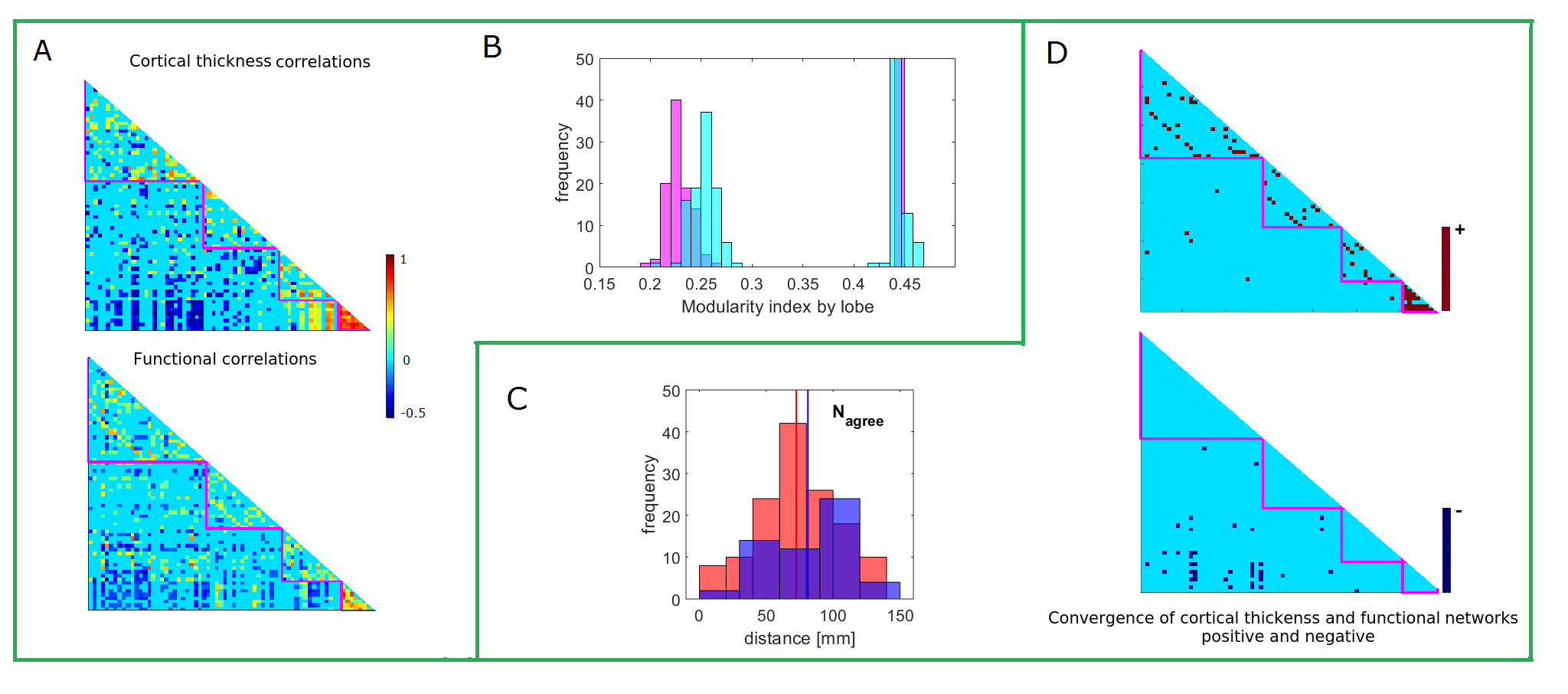}
\caption{Cortical thickness and functional correlations networks. (A) Group-wise cortical thickness (CT) and functional correlations (FC) networks arranged (from top to bottom) according to nodal affiliations with frontal, temporal, parietal and occipital lobes. (B) Histograms of modularity index (Q) by lobe calculated on 100 randomized CT (magenta) and FC (cyan) networks and compared against the modularity index calculated on group-wise CT (magenta) and FC (cyan) networks (values of 0.446 and 0.442, respectively). (C) Histogram of the cortical distances for the convergent positive (red) and negative (blue) networks. Vertical, color-coded lines represent mean value of the corresponding cortical distances (i.e., distances between regions with the overlapping positive or negative correlations). (D) Overlapping edges of positive and negative CT and FC networks; red -- positive correlations; blue -- negative correlations.}
\label{fig:fig_main}
\end{figure}
\subsection{Modularity by Lobes}  
\label{sec:modlob}
Modularity index of the CT and FC networks by lobes were 0.446 and 0.442 (mean values on 100 runs for each network) respectively, thus indicating modular organization of CT/FC networks by cortical lobes. To get the confidence interval for network modularity distribution by lobar (frontal, temporal, parietal and occipital) divisions, 200 randomized networks were generated on CT/FC networks (100 each), i.e., each null network model was tested to obtain distribution of the modularity indexes (Qs) on randomly distributed correlations of the CT/FC network. Fig.~\ref{fig:fig_main}(B) shows how the distributions of 100 Q values on random CT/FC networks differ from those calculated on real networks (i.e., on 100 calculations of the modularity index for each network). 
\subsection{Convergent Cortical Thickness and Functional Networks Maps}
Overlapping thickness and functional networks were calculated by element-wise multiplication of the two matrices. Fig.~\ref{fig:fig_main}(D) shows products of this multiplication for the coupling of positive and negative correlations/networks. The survived network edges of positive correlations map intra-lobar and inter-hemispheric homologous (off-diagonal matrix elements). In contrast, the coupled negative networks map exclusively onto inter-lobar connections. In particular, the frontal lobe regions -- rostral anterior cingulate, rostral middle frontal and superior frontal -- are correlated with regions of the occipital lobe. The mean distance between the coupled networks nodes ($N_{agree}$) was ($72\pm30) mm$ for positive and ($81\pm30) mm$ for negative network nodes ($p>0.05$), Fig.~\ref{fig:fig_main}(C).   
\section{Discussion}
This study demonstrates convergence of the cortical thickness and functional correlations networks, with reference to modular organization of the conventional lobar divisions. Noticeably different patterns of the overlapping correlations were observed for positive and negative networks. In contrast to the positive networks' overlaps, which map almost exclusively within-lobe interactions, negative networks show only between-lobes interactions.

Both cortical networks -- inter-regional thickness and functional correlations -- demonstrated modular organizations that mirror conventional frontal, temporal, parietal and occipital divisions of the cortical surface. Thus allowing for inferences about overlapping interactions of these networks by cortical lobes. While several previous studies reported modular topology of the cortex by cortical morphological (thickness and volume) \cite{chen2008revealing,bassett2008hierarchical} or functional \cite{meunier2009age} measurements, no data are available on functional modules as an instinct property of the cortical surface lobar divisions. Motivated by the lobe-specific patterns of atrophy progression in two types of dementia, we have recently reported similar findings on cortical thickness and surface area networks in healthy elderly subjects \cite{vuk2019cortical}. This study represents first effort to map patterns of positive and negative correlations in cortical thickness and functional networks in healthy adults.

Several interesting results observed when mapping overlapping links between the cortical thickness and functional networks by lobar divisions need further attention. First, the approach adopted here was to compare the networks at a single value of network density.  The advantage of this approach is that is maps only the most pronounced correlations within the both networks. This means that the correlations statistically not different from zero were not considered in the analysis. However, the obtained topologies may include different numbers of correlations for a larger study-group size. Thus, the calculation of the maps of overlapping correlations across a range of networks densities (based also on a larger study group) may reveal some of the cortical interactions that did not pass statistical significance test in this study. Second, the definition of conventional lobar divisions used in this study was based on the Desikan-Kelliany cortical parcellation. Although well established, there exist alternative approaches for the cortical segmentation by using different brain templates \cite{tzourio2002}. Since adoption of different brain templates may have influence on the patterns of network correlations, further studies could validate these results using different anatomical (or functional) classifications. Finally, a study combining structural, functional and diffusion-weighted MRI data could test to what extent variations and strengths of positive and negative correlations can be inferred from underlying direct axonal links or synchronous/asynchronous functional links. These future studies could also resolve problems of estimating networks individually for each subject.

\subsection*{Acknowledgements}
The author would like to acknowledge the support of the Maxwell compute cluster funded by the University of Aberdeen.


\begin{thebibliography}{10}
\expandafter\ifx\csname url\endcsname\relax
  \def\url#1{\texttt{#1}}\fi
\expandafter\ifx\csname urlprefix\endcsname\relax\def\urlprefix{URL }\fi
\providecommand{\bibinfo}[2]{#2}
\providecommand{\eprint}[2][]{\url{#2}}

\bibitem{evans2013networks}
\bibinfo{author}{Evans, A.~C.}
\newblock \bibinfo{title}{Networks of anatomical covariance}.
\newblock \emph{\bibinfo{journal}{Neuroimage}} \textbf{\bibinfo{volume}{80}},
  \bibinfo{pages}{489--504} (\bibinfo{year}{2013}).

\bibitem{lerch2006mapping}
\bibinfo{author}{Lerch, J.~P.} \emph{et~al.}
\newblock \bibinfo{title}{Mapping anatomical correlations across cerebral
  cortex (macacc) using cortical thickness from mri}.
\newblock \emph{\bibinfo{journal}{Neuroimage}} \textbf{\bibinfo{volume}{31}},
  \bibinfo{pages}{993--1003} (\bibinfo{year}{2006}).

\bibitem{gong2012convergence}
\bibinfo{author}{Gong, G.}, \bibinfo{author}{He, Y.}, \bibinfo{author}{Chen,
  Z.~J.} \& \bibinfo{author}{Evans, A.~C.}
\newblock \bibinfo{title}{Convergence and divergence of thickness correlations
  with diffusion connections across the human cerebral cortex}.
\newblock \emph{\bibinfo{journal}{Neuroimage}} \textbf{\bibinfo{volume}{59}},
  \bibinfo{pages}{1239--1248} (\bibinfo{year}{2012}).

\bibitem{hyde2009effects}
\bibinfo{author}{Hyde, K.~L.} \emph{et~al.}
\newblock \bibinfo{title}{The effects of musical training on structural brain
  development: a longitudinal study}.
\newblock \emph{\bibinfo{journal}{Annals of the New York Academy of Sciences}}
  \textbf{\bibinfo{volume}{1169}}, \bibinfo{pages}{182--186}
  (\bibinfo{year}{2009}).

\bibitem{witiuk2014cognitive}
\bibinfo{author}{Witiuk, K.} \emph{et~al.}
\newblock \bibinfo{title}{Cognitive deterioration and functional compensation
  in als measured with fmri using an inhibitory task}.
\newblock \emph{\bibinfo{journal}{Journal of Neuroscience}}
  \textbf{\bibinfo{volume}{34}}, \bibinfo{pages}{14260--14271}
  (\bibinfo{year}{2014}).

\bibitem{vuk2019cortical}
\bibinfo{author}{Vuksnaovi\'c, V.}, \bibinfo{author}{Staff, R.},
  \bibinfo{author}{Ahearn, T.}, \bibinfo{author}{Murray, A.} \&
  \bibinfo{author}{Claude, W.}
\newblock \bibinfo{title}{Cortical thickness and surface area networks in
  healthy aging, alzheimer's disease and behavioral variant fronto-temporal
  dementia}.
\newblock \emph{\bibinfo{journal}{Int J Neur Sys}}  (\bibinfo{year}{in press}).

\bibitem{park2013structural}
\bibinfo{author}{Park, H.-J.} \& \bibinfo{author}{Friston, K.}
\newblock \bibinfo{title}{Structural and functional brain networks: from
  connections to cognition}.
\newblock \emph{\bibinfo{journal}{Science}} \textbf{\bibinfo{volume}{342}},
  \bibinfo{pages}{1238411} (\bibinfo{year}{2013}).

\bibitem{kandel2000principles}
\bibinfo{author}{Kandel, E.~R.} \emph{et~al.}
\newblock \emph{\bibinfo{title}{Principles of neural science}},
  vol.~\bibinfo{volume}{4} (\bibinfo{publisher}{McGraw-hill New York},
  \bibinfo{year}{2000}).

\bibitem{mesulam1998sensation}
\bibinfo{author}{Mesulam, M.-M.}
\newblock \bibinfo{title}{From sensation to cognition.}
\newblock \emph{\bibinfo{journal}{Brain: a journal of neurology}}
  \textbf{\bibinfo{volume}{121}}, \bibinfo{pages}{1013--1052}
  (\bibinfo{year}{1998}).

\bibitem{desikan2006automated}
\bibinfo{author}{Desikan, R.~S.} \emph{et~al.}
\newblock \bibinfo{title}{An automated labeling system for subdividing the
  human cerebral cortex on mri scans into gyral based regions of interest}.
\newblock \emph{\bibinfo{journal}{Neuroimage}} \textbf{\bibinfo{volume}{31}},
  \bibinfo{pages}{968--980} (\bibinfo{year}{2006}).

\bibitem{vuksanovic2014functional}
\bibinfo{author}{Vuksanovi{\'c}, V.} \& \bibinfo{author}{H{\"o}vel, P.}
\newblock \bibinfo{title}{Functional connectivity of distant cortical regions:
  role of remote synchronization and symmetry in interactions}.
\newblock \emph{\bibinfo{journal}{NeuroImage}} \textbf{\bibinfo{volume}{97}},
  \bibinfo{pages}{1--8} (\bibinfo{year}{2014}).

\bibitem{ciric2017benchmarking}
\bibinfo{author}{Ciric, R.} \emph{et~al.}
\newblock \bibinfo{title}{Benchmarking of participant-level confound regression
  strategies for the control of motion artifact in studies of functional
  connectivity}.
\newblock \emph{\bibinfo{journal}{Neuroimage}} \textbf{\bibinfo{volume}{154}},
  \bibinfo{pages}{174--187} (\bibinfo{year}{2017}).

\bibitem{vuksanovic2015dynamic}
\bibinfo{author}{Vuksanovi{\'c}, V.} \& \bibinfo{author}{H{\"o}vel, P.}
\newblock \bibinfo{title}{Dynamic changes in network synchrony reveal
  resting-state functional networks}.
\newblock \emph{\bibinfo{journal}{Chaos: An Interdisciplinary Journal of
  Nonlinear Science}} \textbf{\bibinfo{volume}{25}}, \bibinfo{pages}{023116}
  (\bibinfo{year}{2015}).

\bibitem{WIJ10}
\bibinfo{author}{van Wijk, B.~C.}, \bibinfo{author}{Stam, C.~J.} \&
  \bibinfo{author}{Daffertshofer, A.}
\newblock \bibinfo{title}{Comparing brain networks of different size and
  connectivity density using graph theory}.
\newblock \emph{\bibinfo{journal}{PLoS One}} \textbf{\bibinfo{volume}{5}},
  \bibinfo{pages}{e13701} (\bibinfo{year}{2010}).

\bibitem{RUB10}
\bibinfo{author}{Rubinov, M.} \& \bibinfo{author}{Sporns, O.}
\newblock \bibinfo{title}{Complex network measures of brain connectivity: uses
  and interpretations.}
\newblock \emph{\bibinfo{journal}{NeuroImage}} \textbf{\bibinfo{volume}{52}},
  \bibinfo{pages}{1059--1069} (\bibinfo{year}{2010}).

\bibitem{clauset2004finding}
\bibinfo{author}{Clauset, A.}, \bibinfo{author}{Newman, M.~E.} \&
  \bibinfo{author}{Moore, C.}
\newblock \bibinfo{title}{Finding community structure in very large networks}.
\newblock \emph{\bibinfo{journal}{Physical review E}}
  \textbf{\bibinfo{volume}{70}}, \bibinfo{pages}{066111}
  (\bibinfo{year}{2004}).

\bibitem{chen2008revealing}
\bibinfo{author}{Chen, Z.~J.}, \bibinfo{author}{He, Y.},
  \bibinfo{author}{Rosa-Neto, P.}, \bibinfo{author}{Germann, J.} \&
  \bibinfo{author}{Evans, A.~C.}
\newblock \bibinfo{title}{Revealing modular architecture of human brain
  structural networks by using cortical thickness from mri}.
\newblock \emph{\bibinfo{journal}{Cerebral cortex}}
  \textbf{\bibinfo{volume}{18}}, \bibinfo{pages}{2374--2381}
  (\bibinfo{year}{2008}).

\bibitem{bassett2008hierarchical}
\bibinfo{author}{Bassett, D.~S.} \emph{et~al.}
\newblock \bibinfo{title}{Hierarchical organization of human cortical networks
  in health and schizophrenia}.
\newblock \emph{\bibinfo{journal}{Journal of Neuroscience}}
  \textbf{\bibinfo{volume}{28}}, \bibinfo{pages}{9239--9248}
  (\bibinfo{year}{2008}).

\bibitem{meunier2009age}
\bibinfo{author}{Meunier, D.}, \bibinfo{author}{Achard, S.},
  \bibinfo{author}{Morcom, A.} \& \bibinfo{author}{Bullmore, E.}
\newblock \bibinfo{title}{Age-related changes in modular organization of human
  brain functional networks}.
\newblock \emph{\bibinfo{journal}{Neuroimage}} \textbf{\bibinfo{volume}{44}},
  \bibinfo{pages}{715--723} (\bibinfo{year}{2009}).

\bibitem{tzourio2002}
\bibinfo{author}{Tzourio-Mazoyer, B.} \emph{et~al.}
\newblock \bibinfo{title}{Automated anatomical labeling of activations in {SPM}
  using a macroscopic anatomical parcellation of the {MNI} {MRI} single-subject
  brain}.
\newblock \emph{\bibinfo{journal}{NeuroImage}} \textbf{\bibinfo{volume}{15}},
  \bibinfo{pages}{273--289} (\bibinfo{year}{2002}).

\end{thebibliography}
\end{document}